\documentclass[letterpaper]{article}

\usepackage[english]{babel}
\usepackage[utf8]{inputenc}
\usepackage{amsmath}
\usepackage{graphicx}
\usepackage[labelfont=bf,font=it]{caption}
\usepackage[affil-it]{authblk}
\usepackage[USenglish]{isodate}
\usepackage[]{hyperref}
\usepackage{placeins}

\hypersetup{
    colorlinks=true,
    linkcolor=cyan,
    filecolor=cyan,      
    urlcolor=cyan,
    citecolor=black,
}

\newcommand*\samethanks[1][\value{footnote}]{\footnotemark[#1]}
\newcommand{\subtext}[2]{
#1_{_{\text{#2}}}
}

\title{Which Findings from the Functional Neuroimaging Literature Can We Trust?}

\author{Daniel Kessler\thanks{Corresponding Author: kesslerd@umich.edu} \thanks{Contributed equally}}
\author{Michael Angstadt\samethanks}
\author{Chandra Sripada\samethanks}
\affil{Department of Psychiatry\\ University of Michigan, Ann Arbor}

\date{\today}

\begin{document}
\maketitle

\begin{abstract}
In their recent ``Cluster Failure'' paper, Eklund and colleagues cast doubt on the accuracy of a widely used statistical test in functional neuroimaging. 
Here, we leverage nonparametric methods that control the false discovery rate to offer more nuanced, quantitative guidance about which findings in the existing literature can be trusted. 
We show that, in the task studies examined by Eklund et al., most clusters originally reported to be significant are indeed trustworthy by the false discovery rate benchmark.
\end{abstract}

In a substantial contribution to the functional magnetic resonance imaging (fMRI) field, Eklund et al. \cite{eklund16} use nonparametric methods to demonstrate that random field theory (RFT)-based family-wise error (FWE) correction techniques for cluster-level inference do not control errors as they are supposed to, and this discrepancy is particularly pronounced for lenient cluster defining thresholds (CDT). 
Moreover, they point to violations of RFT assumptions as the culprit for this discrepancy.

Given these results, what advice can we offer to the reader exploring the existing fMRI literature when faced with a table of cluster-wise RFT-based FWE corrected \textit{p}-values ($\subtext{p}{RFT-FWE}$)? 
To suggest caution is reasonable but incomplete; we require concrete, quantitative guidelines to enable appropriate calibration of skepticism.

Here, we undertake an initial attempt to construct such guidance.
We heed Eklund et al.'s warning and prefer null distributions obtained through nonparametric methods rather than through RFT.
However, we focus on the False Discovery Rate (FDR; \cite{fdr}), which is a more natural target for multiple testing control (a point well recognized by Nichols in previous work; \cite{genovese02}).
A researcher determining which clusters are significant is more concerned with the proportion that are false positives (FDR) than whether \textit{any} are false positives (FWE).
Given these considerations, a reader faced with a table of clusters significant under RFT-FWE correction might naturally ask: Which of these results would have survived had the study instead employed a nonparametric FDR-based method?

We address this question using the same task fMRI data \cite{duncan_consistency_2009,tom_neural_2007} analyzed by Eklund et al., which is available from openfMRI \cite{poldrack_toward_2013})  (code, data, and Extended Methods for present analysis are at \url{http://github.com/mangstad/FDR_permutations}).

In brief, for each contrast, we generate 5,000 realizations of the data through sign-flipping. 
To obtain a null distribution of cluster extents (for an arbitrarily chosen cluster), we combine normalized frequency counts of cluster extents calculated at each realization\footnote{Realizations with no clusters assign all their mass to 0.}. 
This distribution is used to assign uncorrected \textit{p}-values to each observed cluster. 
We next submit the vector of uncorrected \textit{p}-values for each contrast to Benjamini and Hochberg's \cite{fdr} FDR procedure with $\subtext{\alpha}{FDR}=.05$.
 
We compare $\subtext{p}{RFT-FWE}$-values to $\subtext{p}{FDR}$-values and note whether they remain significant under $\subtext{\alpha}{FDR}=.05$. 
We generate separate plots for this analysis conducted at CDT=\{.001, .01\}.

\begin{figure}[ht]
\includegraphics[width=1.0\textwidth]{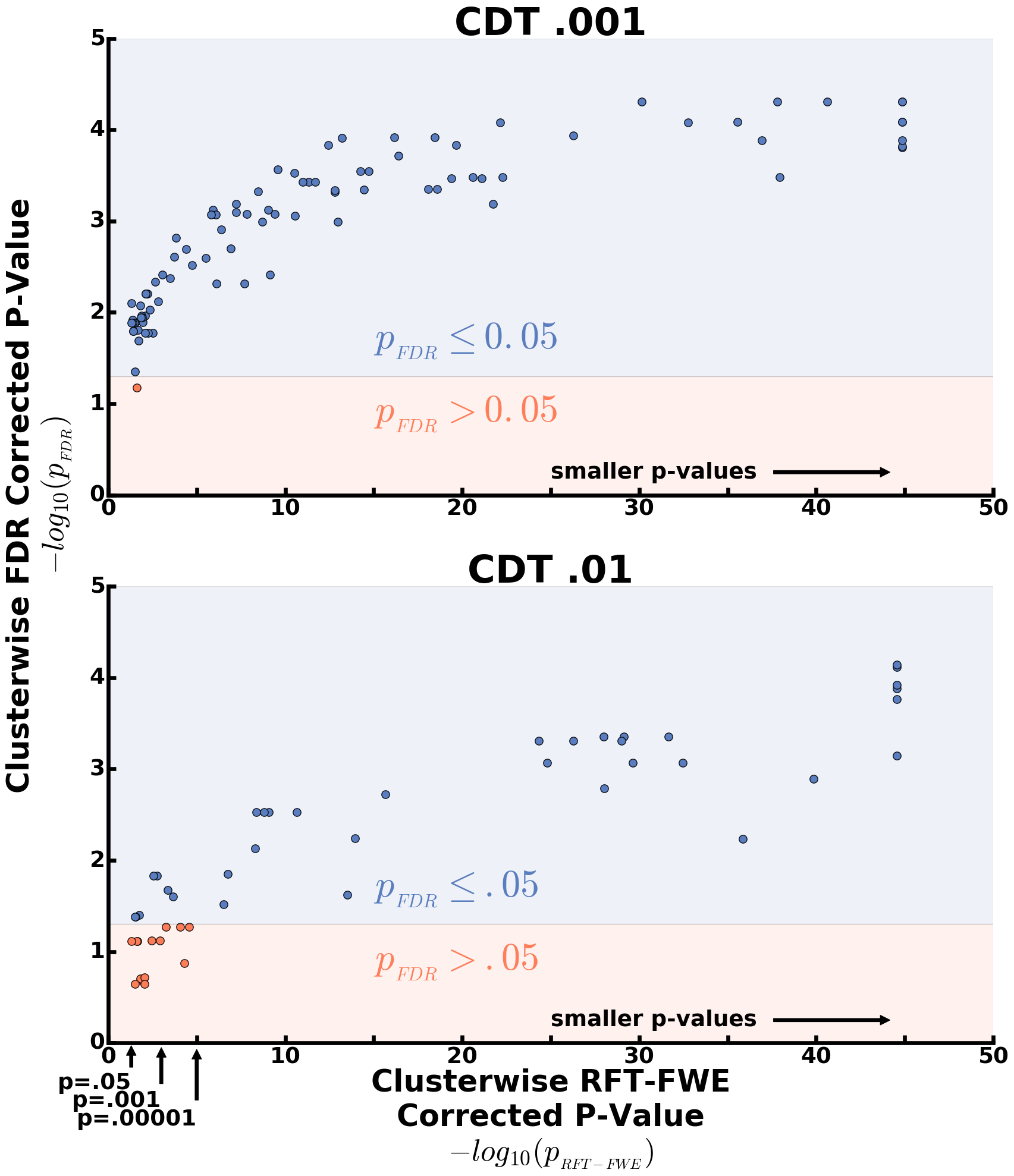}
\centering
\caption{
\textbf{Assessing RFT-Based FWE Using an FDR Benchmark.}
We submitted the same task data analyzed by Eklund et al. to nonparametric cluterwise FDR analysis. For $\text{CDT}=.001$ (top), RFT-based FWE approximates effective FDR control with $\subtext{\alpha}{FDR} = .05$. For $\text{CDT}=.01$ (bottom), only clusters with $\subtext{p}{RFT-FWE} \leq .00001$ were reliably significant at $\subtext{\alpha}{FDR}=.05$. 
\label{fig:p-plot}
}
\end{figure} 

Results (see Figure \ref{fig:p-plot}) show that for CDT=.001, only one cluster significant at $\subtext{\alpha}{RFT-FWE}\leq .05$ failed to be significant at $\subtext{\alpha}{FDR}\leq .05$, thus suggesting RFT-based FWE closely approximates effective FDR control.
This finding has promising implications for the existing body of fMRI studies using RFT-based cluster-level inference that used this stricter CDT, estimated to be upwards of 8,500 reports \cite{nichols_blog,woo14}.
For CDT=.01, used in approximately 3,500 studies \cite{nichols_blog,woo14}, Eklund et al. and others \cite{flandin_analysis_2016} have urged caution and our results agree. We found that $\subtext{\alpha}{RFT-FWE}$ must be very strict (at least .00001) for effective FDR control to be achieved (i.e., many $.05 \geq \subtext{p}{RFT-FWE} \geq .00001$ fail to meet significance at $\subtext{\alpha}{FDR}=.05$).

These results offer initial quantitative guidance on interpreting the past literature that employed RFT-based FWE, providing a more granular appreciation of the relationship between $\subtext{p}{RFT-FWE}$ and trustworthiness of the result.
A more comprehensive examination of fMRI task data sets that used RFT-based FWE may further refine this guidance.

\FloatBarrier
\bibliography{biblio}
\bibliographystyle{plain}

\end{document}